\documentclass[final,5p,times,twocolumn]{elsarticle}
\usepackage{lineno,hyperref,amsmath,amsthm,amssymb,amsfonts,ragged2e,color,subfig}
\journal{``High Energy Density Physics"}
\begin{document}
\begin{frontmatter}
\title{Ion-acoustic rogue waves in multi-ion plasmas}
\author{M. Hassan$^{*,1}$, M. H. Rahman$^{1}$, N. A. Chowdhury$^{**,1}$, A. Mannan$^{1,3}$, and A. A. Mamun$^{1,2}$}
\address{$^{1}$Department of Physics, Jahangirnagar University, Savar, Dhaka-1342, Bangladesh\\
$^{2}$Wazed Miah Science Research Center, Jahangirnagar University, Savar, Dhaka-1342, Bangladesh\\
$^{3}$Institut f\"{u}r Mathematik, Martin Luther Universit\"{a}t Halle-Wittenberg, Halle, Germany\\
Email: $^*$hassan206phy@gmail.com, $^{**}$nurealam1743phy@gmail.com}
\begin{abstract}
The basic properties of nonlinear ion-acoustic (IA) waves (IAWs), particularly finite amplitude IA rogue
waves (IARWs) in a plasma medium (containing pair ions, iso-thermal positrons and non-thermal electrons) are
theoretically investigated by deriving the nonlinear Schr\"{o}dinger equation (NLSE). The criteria for the
modulational instability of IAWs, and the basic features of finite amplitude IARWs are identified. The modulationally
stable and unstable regions are determined by the sign of the ratio of the dispersive coefficient to the nonlinear
coefficient of NLSE. The latter is analyzed to obtain the region for the existence of the IARWs, which corresponds
to the unstable region. The shape of the profile of the rogue waves depends on the non-thermal parameter
$\alpha$ and the ratio of electron temperature to positron temperature. It is found that the increase in the
value of the non-thermal parameter enhances both the amplitude and width of IARWs, and that the enhancement of
positron (electron) temperature reduces (enhances) the amplitude and width of IARWs. It is worth to mention
that our present investigation may be useful for understanding the salient features of IARWs in
space (viz., upper region of Titan's atmosphere, cometary comae, and Earth's ionosphere, etc.) and laboratory (viz., plasma processing
reactor and neutral beam sources, etc.) plasmas.
\end{abstract}
\end{frontmatter}
\section{Introduction}
\label{1sec:int}
The signature of pair-ion (PI) plasma (PIP) has been observed by the Cassini spacecraft \cite{Coates2007}
in spaces, specially, in upper region of Titan's atmosphere, and is also generated
by Oohara \textit{et al.} \cite{Oohara2003,Oohara2005,Oohara2007} in laboratory experiments.
PI has also identified in cometary comae \cite{Sabry2009,Jannat2015,Jannat2016},
Earth's ionosphere \cite{El-Labany2012,Massey1976,El-Tantawy2013,Abdelwahed2016,Elwakil2010},
plasma processing reactor \cite{Gottscho1986}, neutral beam sources \cite{Bacal1979},
and successively has encouraged many researchers to understand the intrinsic contribution
of the PI in recognizing the collective properties of the PIP by investigating ion-plasma waves \cite{Oohara2005},
intermediate frequency waves \cite{Oohara2005}, ion-acoustic (IA) solitary waves (IASWs) \cite{Sabry2009},
IA shock waves (IASHWs) \cite{Jannat2015}, IA Gardner solitons (IAGSs) \cite{Jannat2016},
IA double layers (IADLs) \cite{El-Labany2012}, and IA rogue waves (IARWs) \cite{El-Tantawy2013,Abdelwahed2016,Elwakil2010,C1,C7}, etc.

Non-thermal particles are appeared in a complex system due to the presence of the external force
field, and are governed by the Cairns distribution/non-thermal distribution \cite{Cairns1995,C4,C6}, and have also been
identified by the Freja \cite{Dovner1994} and Viking \cite{Bostrom1988} satellites in the magnetosphere and upper ionosphere of the auroral zone.
Paul and Bandyopadhyay  \cite{Paul2016} investigated dust-ion-acoustic solitary waves in a four components plasma medium having non-thermal plasma species,
and observed that the height of the positive super-solitons increases with increasing the value of non-thermality of the plasma species.
Selim \cite{Selim2016} studied IASWs in a three components PIP medium (PIPM) by considering non-thermal electrons, and
found that the height and thickness of the compressive solitons increase with increasing the value of electron's non-thermality.
Singh and Lakhina \cite{Singh2015} considered a three components plasma model having non-thermal electrons, and studied IA super-solitons,
and found that the width of the super-solitons increases with an increase in the value of electrons non-thermality.

Electrostatic rogue waves (RWs) are considered to have appeared due to
the nonlinear property of the plasma medium as well as the modulational instability (MI) of
the carrier waves, and are governed by the nonlinear Schr\"{o}dinger equation (NLSE),
and have also been identified in the plasma physics \cite{El-Tantawy2013,Abdelwahed2016,Elwakil2010,Rahman2018}.
El-Tantawy \textit{et al.} \cite{El-Tantawy2013} considered a two components plasma model
having inertial ions and non-thermal electrons, and studied the mechanism of the formation of
IARWs, and also found that behind the critical value of the non-thermal parameter, the nonlinearity
and height of the IARWs increase with the non-thermal parameter. Abdelwahed \textit{et al.} \cite{Abdelwahed2016}
considered a three components PIPM and observed the IARWs in presence of super-thermal electrons, and found that the height of the IARWs
decreases with the positive ion number density. Elwakil \textit{et al.} \cite{Elwakil2010} analyzed the MI of the IAWs in a
PIPM having non-thermal electrons. Rahman \textit{et al.} \cite{Rahman2018} studied dust-acoustic RWs in a non-thermal dusty plasma
medium, and found that the height of the RWs increases with the non-thermality of the plasma species.

Recently, Sabry \textit{et al.} \cite{Sabry2009} studied IASWs in a three components PIPM having non-thermal electrons, and
found that the presence of negative ion enhances the amplitude but decreases the width of the solitary pulse.
El-Labany \textit{et al.} \cite{El-Labany2012} reported IADLs in a three components
PIPM in presence of  non-thermal electrons. Jannat \textit{et al.} \cite{Jannat2015} examined IASHWs in a
four components PIPM having inertial PI as well as inertialess electrons and positrons.
It is inevitable to investigate the MI of the IAWs and the formation of the IARWs in a four
components PIPM by considering inertial double ions and inertialess iso-thermal positrons
as well as non-thermal Cairns distributed electrons.

The outline of the paper is as follows: The governing equations describing our plasma
model are presented in Section \ref{1sec:Governing Equation}. The standard NLSE is derived in Section \ref{1sec:Derivation of the NLSE}.
The MI of the IAWs is examined in Section \ref{1sec:Modulational instability}. The first-order and
second-order RWs are discussed in Section \ref{1sec:Rogue waves}. Finally, a brief conclusion is provided in Section \ref{1sec:Conclusion}.
\section{Governing Equations}
\label{1sec:Governing Equation}
We consider a four components plasma model consisting of negative ions, positive ions, electrons, and positrons.
Overall, the charge neutrality condition can be written as $Z_+ n_{+0}+n_{p0}=Z_-n_{-0}+n_{e0}$; where $n_{+0}$,
$n_{p0}$, $n_{-0}$, and $n_{e0}$ are, respectively, the equilibrium number densities of positive ions, iso-thermal positrons,
negative ions, and non-thermal Cairn's distributed electrons, and also $Z_+$ and $Z_-$ are, respectively, the charge state of
the positive ions and negative ions. Now, the basic set of normalized equations can be written as
\begin{eqnarray}
&&\hspace*{-1.3cm}\frac{\partial n_+}{\partial t}+\frac{\partial}{\partial x}(n_+u_+)=0,
\label{1eq:1}\\
&&\hspace*{-1.3cm}\frac{\partial u_+}{\partial t}+u_+\frac{\partial u_+}{\partial x}=-\frac{\partial\phi}{\partial x},
\label{1eq:2}\\
&&\hspace*{-1.3cm}\frac{\partial n_-}{\partial t}+\frac{\partial}{\partial x}(n_-u_-)=0,
\label{1eq:3}\\
&&\hspace*{-1.3cm}\frac{\partial u_-}{\partial t}+u_-\frac{\partial u_-}{\partial x}=\delta_1\frac{\partial\phi}{\partial x},
\label{1eq:4}\\
&&\hspace*{-1.3cm}\frac{\partial^2\phi}{\partial x^2}=\delta_2 n_e-\delta_3n_p+(1-\delta_2+\delta_3)n_--n_+,
\label{1eq:5}
\end{eqnarray}
where $n_+$ and $n_-$  are the positive and negative ion number density normalized
by their equilibrium value $n_{+0}$ and $n_{-0}$, respectively; $u_-$ and $u_+$ are
the negative and positive ion fluid speed normalized by wave speed $C_+=(Z_+k_BT_e/m_+)^{1/2}$
(with $T_e$ being the temperature non-thermal electron, $m_+$ being the positive ion mass,
and $k_B$ being the Boltzmann constant); $\phi$ is the electrostatic wave potential
normalized by $k_BT_e/e$ (with $e$ being the magnitude of single electron charge); the
time and space variables are normalized by $\omega_{p+}^{-1}=(m_+/4\pi Z_+^2 e^2 n_{+0})^{1/2}$
and $\lambda_{D+}=(k_BT_e/4\pi e^2 Z_+n_{+0})^{1/2}$, respectively. Other plasma parameters are defined as $\delta_1=Z_-m_+/Z_+m_-$,
$\delta_2=n_{e0}/Z_+n_{+0}$, and $\delta_3=n_{p0}/Z_+n_{+0}$. Now, the expression for  electron number
density which is obeying non-thermal Cairn's distribution \cite{Cairns1995} is given by
\begin{eqnarray}
&&\hspace*{-1.3cm}n_e=(1-\beta\phi+\beta\phi^2)\exp(\phi),
\label{1eq:6}
\end{eqnarray}
where $\beta=4\alpha/(1+3\alpha)$, with $\alpha$ being the parameter determining the fast particles present in our plasma
model. The positron number density which is iso-thermally distributed is given by
\begin{eqnarray}
&&\hspace*{-1.3cm}n_p=\exp(-\delta_4\phi),
\label{1eq:7}
\end{eqnarray}
where $\delta_4=T_e/T_p$ and $T_e>T_p$. Now, by substituting Eq. \eqref{1eq:6} and \eqref{1eq:7} into Eq. \eqref{1eq:5},
and expanding up to third order of $\phi$, we get
\begin{eqnarray}
&&\hspace*{-1.3cm}\frac{\partial^2\phi}{\partial x^2}+n_+=(\delta_2-\delta_3)+(1-\delta_2+\delta_3)n_-
\nonumber\\
&&\hspace*{0.3cm}+H_1\phi+ H_2 \phi^2+ H_3\phi^3+\cdot\cdot\cdot,
\label{1eq:8}
\end{eqnarray}
where
\begin{eqnarray}
&&\hspace*{-1.3cm}H_1=\delta_2(1-\beta)+\delta_3\delta_4,
\nonumber\\
&&\hspace*{-1.3cm}H_2=[\delta_2-\delta_3\delta_4^2]/2,
\nonumber\\
&&\hspace*{-1.3cm}H_3=[\delta_2(1-\beta+6\beta)+\delta_3\delta_4^3]/6.
\nonumber\
\end{eqnarray}
We note that the term on the right hand side of the Eq. \eqref{1eq:8}  is the contribution of electron and positron species.
\section{Derivation of the NLSE}
\label{1sec:Derivation of the NLSE}
To study the MI of IAWs, we want to derive the NLSE by employing the reductive perturbation method (RPM)
and for that case, we can write the stretched coordinates in the form \cite{Kourakis2003,Kourakis2005}
\begin{eqnarray}
&&\hspace*{-1.3cm}\xi={\epsilon}(x-v_g t),
\label{1eq:9}\\
&&\hspace*{-1.3cm}\tau={\epsilon}^2 t,
\label{1eq:10}
\end{eqnarray}
where $v_g$ is the group velocity and $\epsilon$ ($\epsilon\ll 1$) is a small parameter. Then, we can write the dependent variables  as \cite{Kourakis2003,Kourakis2005}
\begin{eqnarray}
&&\hspace*{-1.3cm}n_+=1+\sum_{m=1}^{\infty}\epsilon^{m}\sum_{l=-\infty}^{\infty} n_{+l}^{(m)}(\xi,\tau)~\mbox{exp}[i l(kx-\omega t)],
\label{1eq:11}\\
&&\hspace*{-1.3cm}u_+=\sum_{m=1}^{\infty}\epsilon^{m}\sum_{l=-\infty}^{\infty} u_{+l}^{(m)}(\xi,\tau)~\mbox{exp}[i l(kx-\omega t)],
\label{1eq:12}\\
&&\hspace*{-1.3cm}n_-=1+\sum_{m=1}^{\infty}\epsilon^{m}\sum_{l=-\infty}^{\infty} n_{-l}^{(m)}(\xi,\tau)~\mbox{exp}[i l(kx-\omega t)],
\label{1eq:13}\\
&&\hspace*{-1.3cm}u_-=\sum_{m=1}^{\infty}\epsilon^{m}\sum_{l=-\infty}^{\infty} u_{-l}^{(m)}(\xi,\tau)~\mbox{exp}[i l(kx-\omega t)],
\label{1eq:14}\\
&&\hspace*{-1.3cm}\phi=\sum_{m=1}^{\infty}\epsilon^{m}\sum_{l=-\infty}^{\infty} \phi_{l}^{(m)}(\xi,\tau)~\mbox{exp}[i l(kx-\omega t)],
\label{1eq:15}\
\end{eqnarray}
where $k$ ($\omega$) is real variables representing the carrier wave number (frequency). The derivative operators in above
equations are treated as follows:
\begin{eqnarray}
&&\hspace*{-1.3cm}\frac{\partial}{\partial t}\rightarrow\frac{\partial}{\partial t}-\epsilon v_g \frac{\partial}{\partial\xi}
+\epsilon^2\frac{\partial}{\partial\tau},
\label{1eq:16}\\
&&\hspace*{-1.3cm}\frac{\partial}{\partial x}\rightarrow\frac{\partial}{\partial x}+\epsilon\frac{\partial}{\partial\xi}.
\label{1eq:17}
\end{eqnarray}
Now, by substituting Eq. \eqref{1eq:9}-\eqref{1eq:17}  into  Eq. \eqref{1eq:1}-\eqref{1eq:4} and Eq. \eqref{1eq:8}, and
collecting the terms containing $\epsilon$, the first order ($m=1$ with $l=1$)  reduced equations can be written as
\begin{eqnarray}
&&\hspace*{-1.3cm}n_{+1}^{(1)}=\frac{k^2}{\omega^2}\phi_1^{(1)},
\label{1eq:18}\\
&&\hspace*{-1.3cm}u_{+1}^{(1)}=\frac{k}{\omega}\phi_1^{(1)},
\label{1eq:19}\\
&&\hspace*{-1.3cm}n_{-1}^{(1)}=-\frac{k^2\delta_1}{\omega^2}\phi_1^{(1)},
\label{1eq:20}\\
&&\hspace*{-1.3cm}u_{-1}^{(1)}=-\frac{k\delta_1}{\omega}\phi_1^{(1)}.
\label{1eq:21}\
\end{eqnarray}
These relation provides the dispersion relation for IAWs
\begin{eqnarray}
&&\hspace*{-1.3cm}\omega^2=\frac{k^2[1+\delta_1(1-\delta_2+\delta_3)]}{k^2+H_1}.
\label{1eq:22}
\end{eqnarray}
%%%%%%%%%%%%%%%%%%%%%%%%%%%%%%
\begin{figure}
\centering
\includegraphics[width=85mm]{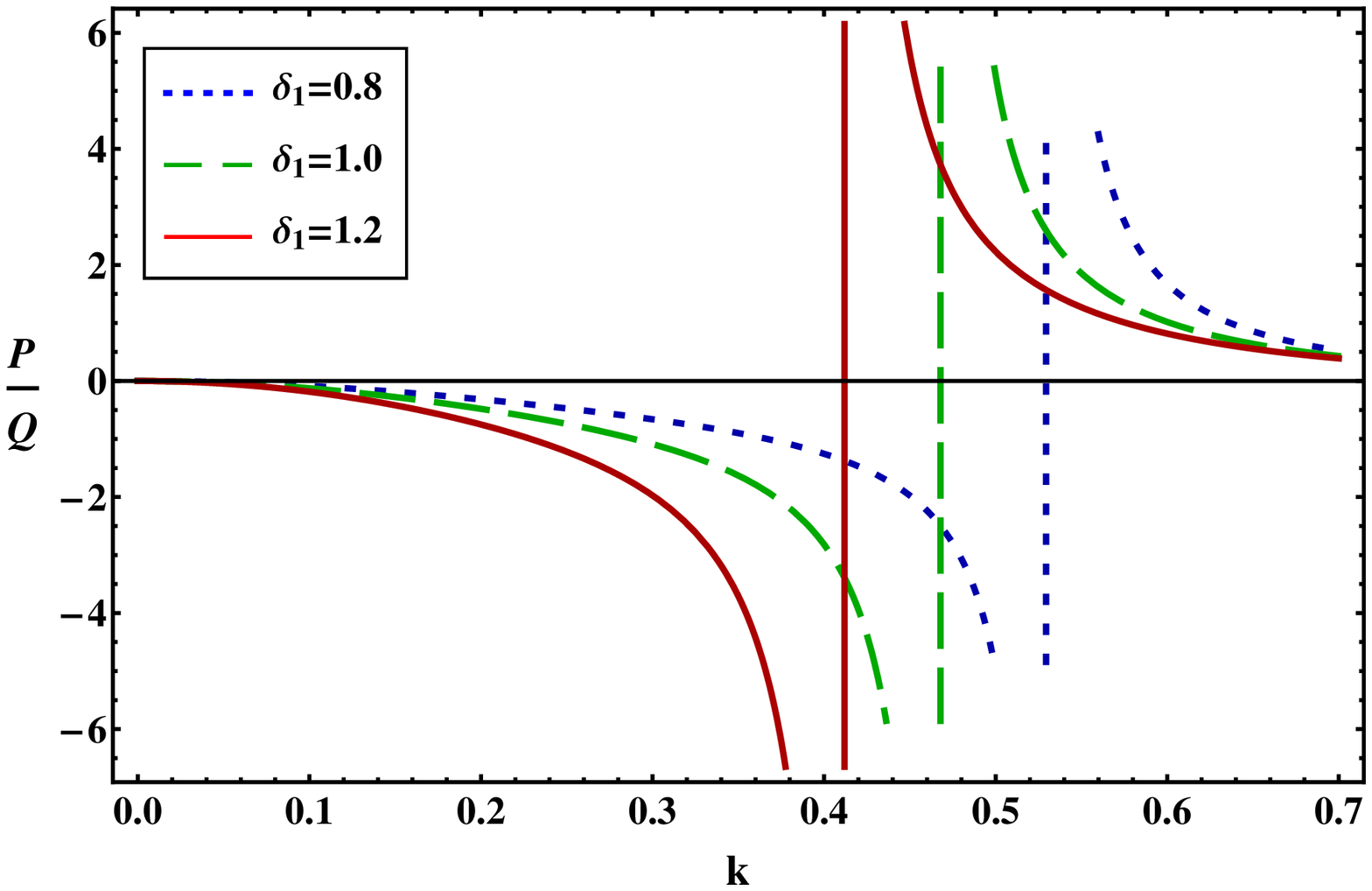}
\caption{Plot of $P/Q$ vs $k$ for different values of $\delta_1$ when other parameters
are $\alpha=0.7$, $\delta_2=0.4$, $\delta_3=0.4$, and $\delta_4=2.0$.}
\label{1Fig:F1}
\end{figure}
\begin{figure}
\centering
\includegraphics[width=85mm]{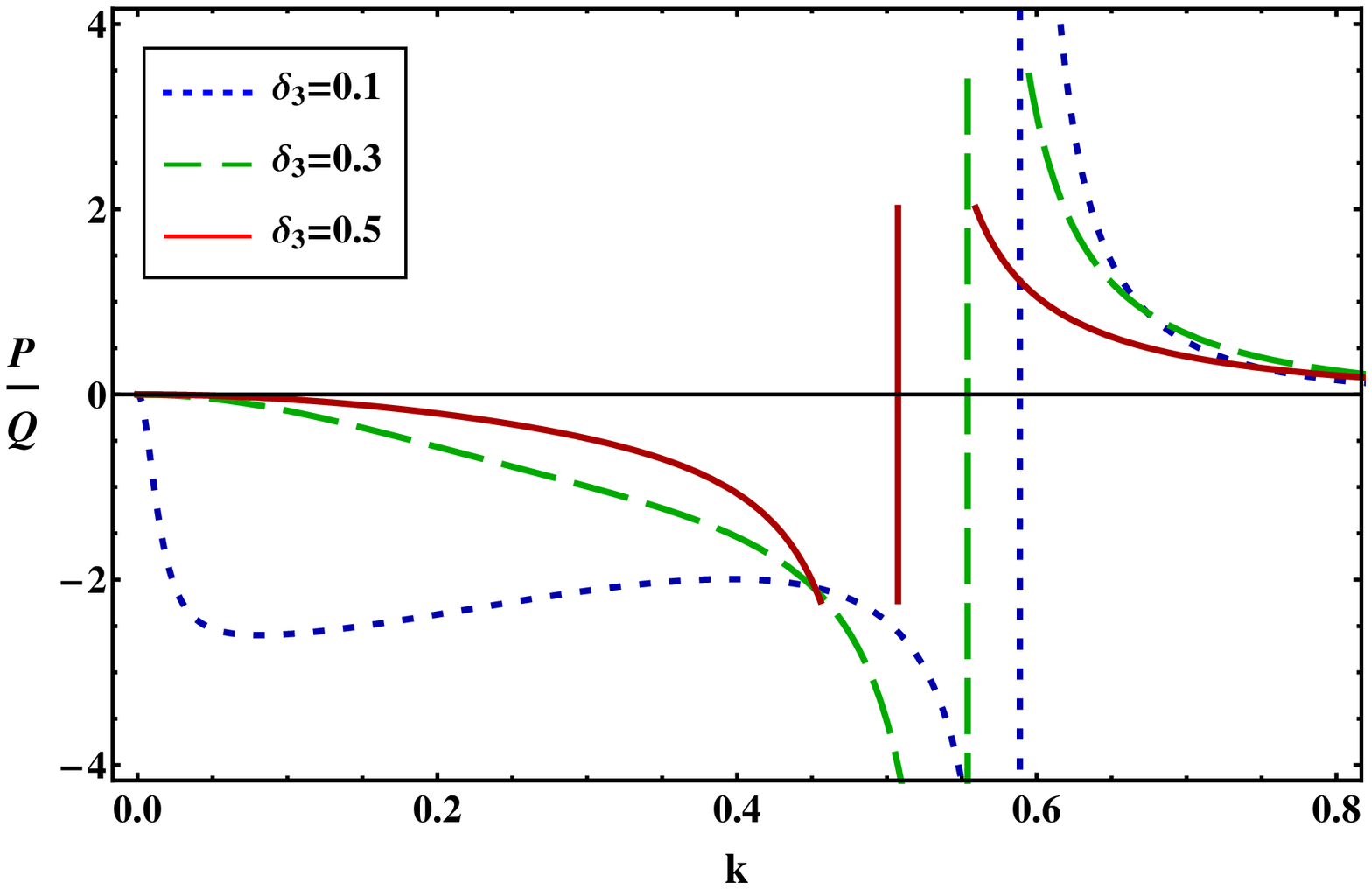}
\caption{Plot of $P/Q$ vs $k$ for different values of $\delta_3$ when other parameters
are $\alpha=0.7$, $\delta_1=0.8$, $\delta_2=0.4$, and $\delta_4=2.0$.}
\label{1Fig:F2}
\end{figure}
%%%%%%%%%%%%%%%%%%%%%%%%%%
%%%%%%%%%%%%%%%%%%%%%%%%%%%%%%%%%%%%%%%%%%%%%%%%%%%%%%%%%%%%%%
\begin{figure}
\centering
\includegraphics[width=85mm]{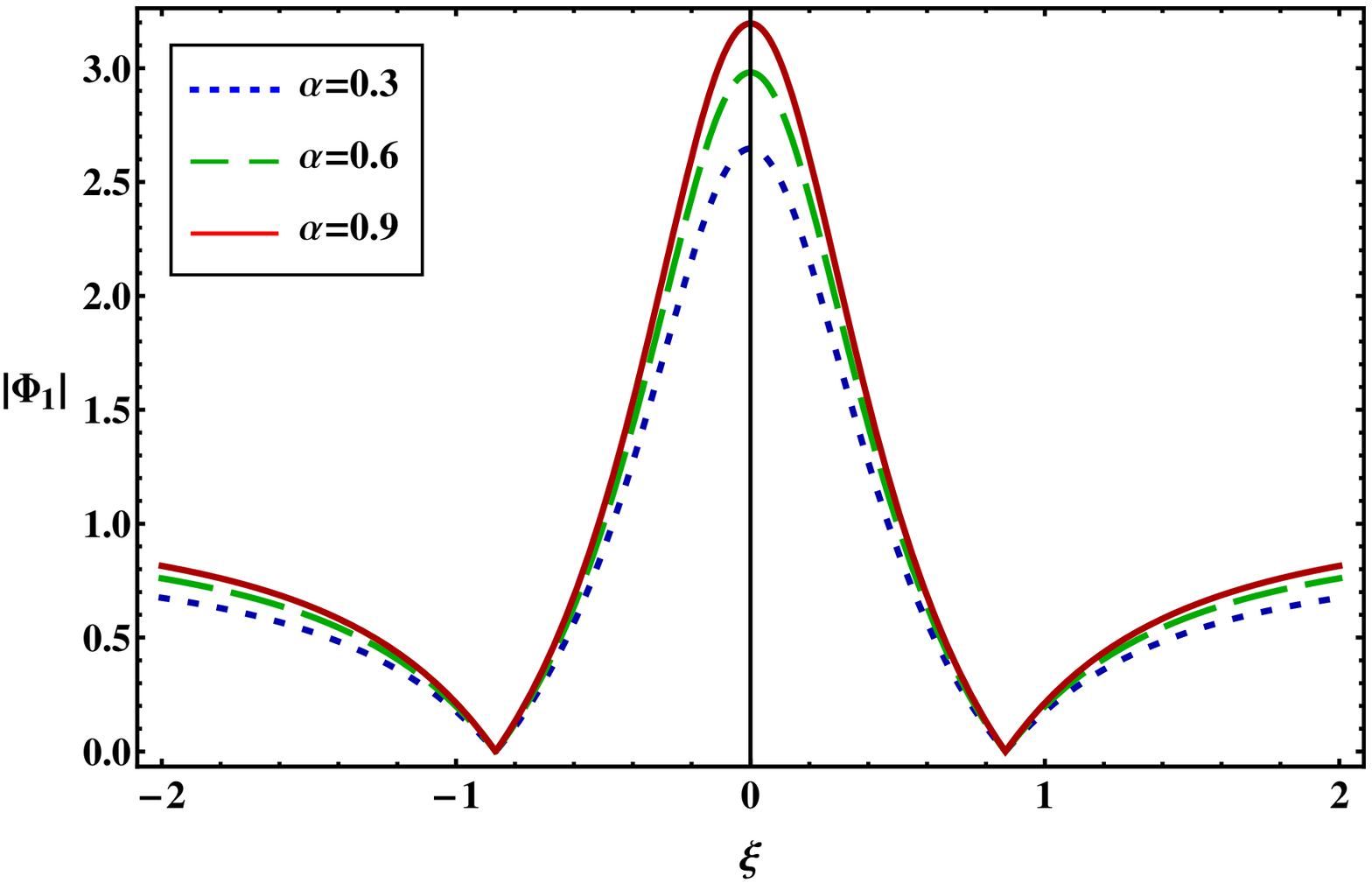}
\caption{Plot of $|\Phi_1|$ vs $\xi$ for different values of $\alpha$ when other parameters
are $k=0.7$, $\tau=0$, $\delta_1=0.8$, $\delta_2=0.4$, $\delta_3=0.4$, and $\delta_4=2.0$.}
\label{1Fig:F3}
\end{figure}
\begin{figure}
\centering
\includegraphics[width=85mm]{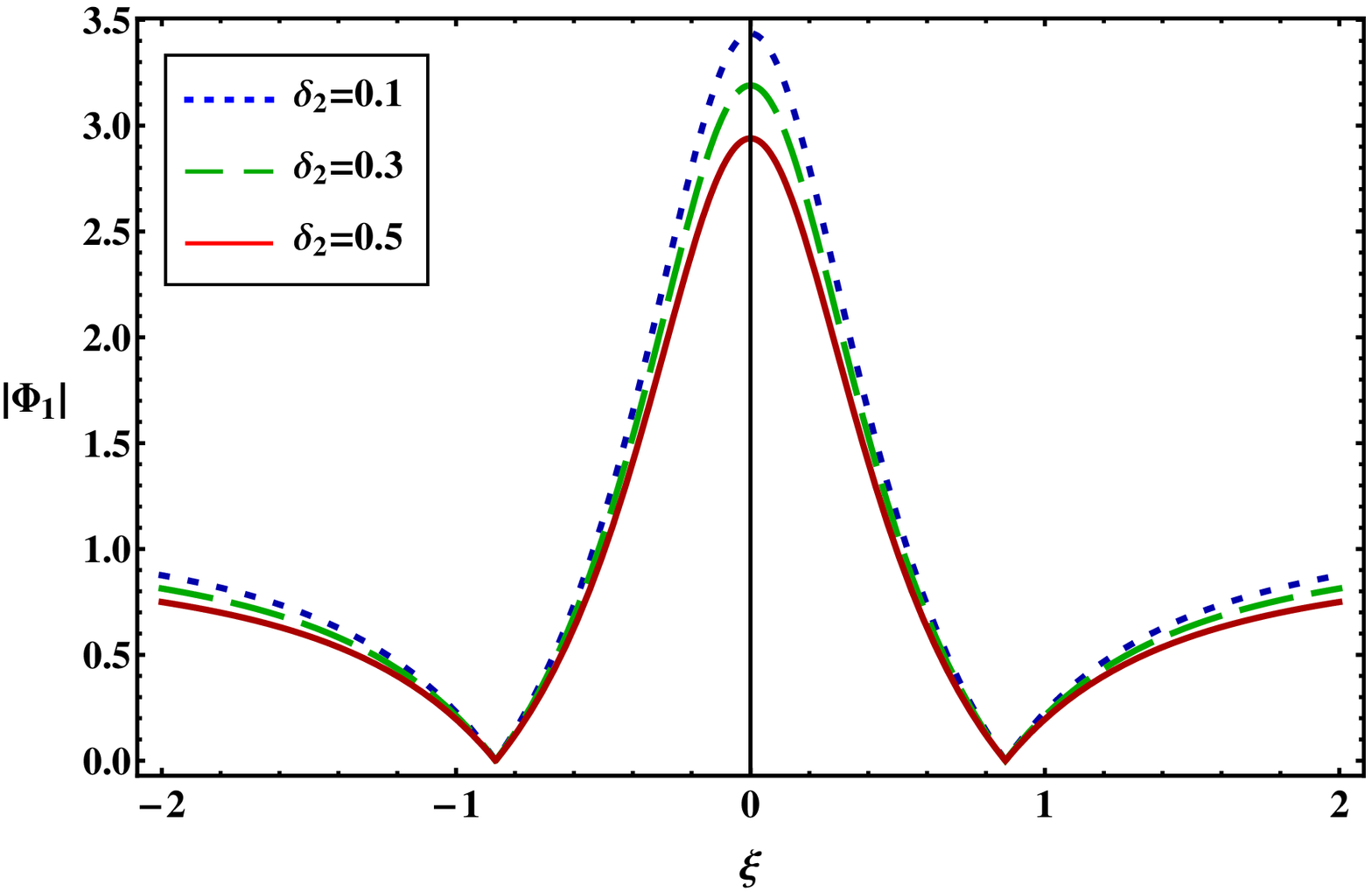}
\caption{Plot of $|\Phi_1|$ vs $\xi$ for different values of $\delta_2$ when other parameters
are $k=0.7$, $\tau=0$, $\alpha=0.7$, $\delta_1=0.8$, $\delta_3=0.4$, and $\delta_4=2.0$.}
\label{1Fig:F4}
\end{figure}
\begin{figure}
\centering
\includegraphics[width=85mm]{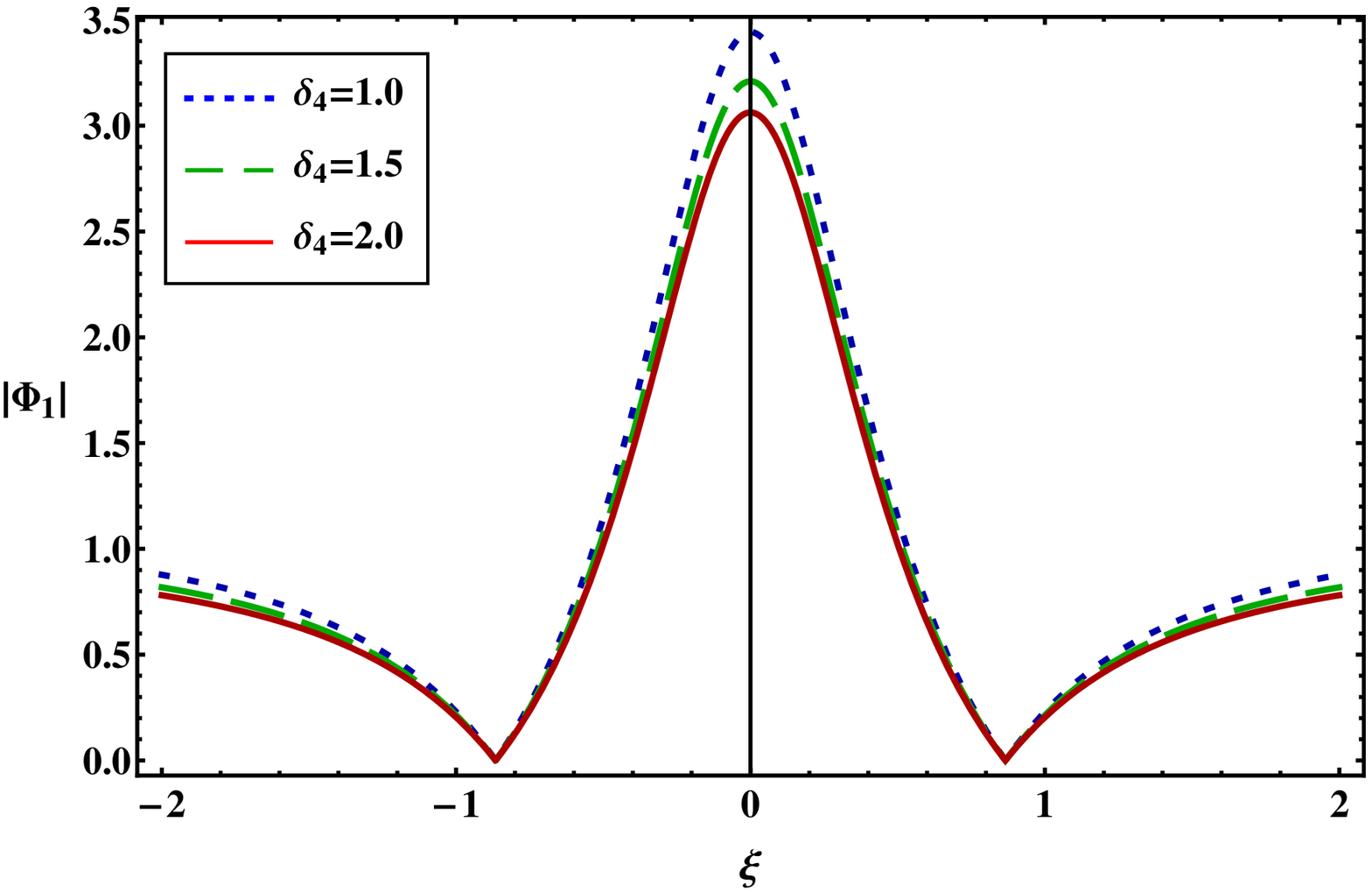}
\caption{Plot of $|\Phi_1|$ vs $\xi$ for different values of $\delta_4$ when other parameters
are $k=0.7$, $\tau=0$, $\alpha=0.7$, $\delta_1=0.8$, $\delta_2=0.4$, and $\delta_3=0.4$.}
\label{1Fig:F5}
\end{figure}
%%%%%%%%%%%%%%%%%%%%%%%%%%%%%%%%%%%%%%%%%%%%%%%%%%%%%%%%%%%%
The second order ($m=2$ with $l=1$) equations are given by
\begin{eqnarray}
&&\hspace*{-1.3cm}n_{+1}^{(1)}=\frac{k^2}{\omega^2}\phi_1^{(2)}+\frac{2ik(v_gk-\omega)}{\omega^3}\frac{\partial\phi_1^{(1)}}{\partial\xi},
\label{1eq:23}\\
&&\hspace*{-1.3cm}u_{+1}^{(2)}=\frac{k}{\omega}\phi_1^{(2)}+\frac{i(v_gk-\omega)}{\omega^2}\frac{\partial\phi_1^{(1)}}{\partial\xi},
\label{1eq:24}\\
&&\hspace*{-1.3cm}n_{-1}^{(2)}=-\frac{\delta_1k^2}{\omega^2}\phi_1^{(2)}-\frac{2ik\delta_1(v_gk-\omega)}{\omega^3}\frac{\partial\phi_1^{(1)}}{\partial\xi},
\label{1eq:25}\\
&&\hspace*{-1.3cm}u_{-1}^{(2)}=-\frac{\delta_1 k}{\omega}\phi_1^{(2)}-\frac{i\delta_1(v_gk-\omega)}{\omega^2}\frac{\partial\phi_1^{(1)}}{\partial\xi},
\label{1eq:26}\
\end{eqnarray}
with the compatibility condition
\begin{eqnarray}
&&\hspace*{-1.3cm}v_g=\frac{\omega[1+\delta_1(1-\delta_2+\delta_3)-\omega^2]}{k[1+\delta_1(1-\delta_2+\delta_3)]}.
\label{1eq:27}
\end{eqnarray}
The coefficients of $\epsilon$ for $m=2$ and $l=2$ provide the second order harmonic amplitudes which are found
to be proportional to $|\phi_1^{(1)}|^2$
\begin{eqnarray}
&&\hspace*{-1.3cm}n_{+2}^{(2)}=H_4|\phi_1^{(1)}|^2,
\label{1eq:28}\\
&&\hspace*{-1.3cm}u_{+2}^{(2)}=H_5 |\phi_1^{(1)}|^2,
\label{1eq:29}\\
&&\hspace*{-1.3cm}n_{-2}^{(2)}=H_6|\phi_1^{(1)}|^2,
\label{1eq:30}\\
&&\hspace*{-1.3cm}u_{-2}^{(2)}=H_7 |\phi_1^{(1)}|^2,
\label{1eq:31}\\
&&\hspace*{-1.3cm}\phi_{2}^{(2)}=H_8 |\phi_1^{(1)}|^2,
\label{1eq:32}\
\end{eqnarray}
where
\begin{eqnarray}
&&\hspace*{-1.3cm}H_4=\frac{3k^4+2H_8\omega^2k^2}{2\omega^4},
\nonumber\\
&&\hspace*{-1.3cm}H_5=\frac{k^3+2\omega^2kH_8}{2\omega^3},
\nonumber\\
&&\hspace*{-1.3cm}H_6=\frac{3\delta_1^2k^4-2\delta_1H_8\omega^2k^2}{2\omega^4},
\nonumber\\
&&\hspace*{-1.3cm}H_7=\frac{\delta_1^2k^3-2\omega^2k\delta_1H_8}{2\omega^3},
\nonumber\\
&&\hspace*{-1.3cm}H_8=\frac{2H_2\omega^4+3\delta_1k^4(1-\delta_2+\delta_3)-3k^4}{2\omega^2k^2-2\omega^4(4k^2+H_1)+2\delta_1\omega^2k^2(1-\delta_2+\delta_3)}.
\nonumber\
\end{eqnarray}
Now, we consider the expression for ($m=3$ with $l=0$) and ($m=2$ with $l=0$), which leads zeroth harmonic modes.
Thus, we obtain
\begin{eqnarray}
&&\hspace*{-1.3cm}n_{+0}^{(2)}=H_{9}|\phi_1^{(1)}|^2,
\label{1eq:33}\\
&&\hspace*{-1.3cm}u_{+0}^{(2)}=H_{10}|\phi_1^{(1)}|^2,
\label{1eq:34}\\
&&\hspace*{-1.3cm}n_{-0}^{(2)}=H_{11}|\phi_1^{(1)}|^2,
\label{1eq:35}\\
&&\hspace*{-1.3cm}u_{-0}^{(2)}=H_{12}|\phi_1^{(1)}|^2,
\label{1eq:36}\\
&&\hspace*{-1.3cm}\phi_0^{(2)}=H_{13} |\phi_1^{(1)}|^2,
\label{1eq:37}\
\end{eqnarray}
where
\begin{eqnarray}
&&\hspace*{-1.3cm}H_{9}=\frac{2v_gk^3+\omega k^2+\omega^3H_{13}}{v_g^2\omega^3},
\nonumber\\
&&\hspace*{-1.3cm}H_{10}=\frac{k^2+\omega^2 H_{13}}{v_g\omega^2},
\nonumber\\
&&\hspace*{-1.3cm}H_{11}=\frac{2v_g\delta_1^2k^3+\omega\delta_1^2k^2-\delta_1\omega^3H_{13}}{v_g^2\omega^3},
\nonumber\\
&&\hspace*{-1.3cm}H_{12}=\frac{\delta_1^2k^2-\delta_1\omega^2H_{13}}{v_g\omega^2},
\nonumber\\
&&\hspace*{-1.3cm}H_{13}=\frac{2H_2v_g^2\omega^3+2v_g\delta_1^2k^3(1-\delta_2+\delta_3)+F1}{\omega^3 [1+\delta_1(1-\delta_2+\delta_3)-H_1v_g^2]},
\nonumber\
\end{eqnarray}
where $F1=\omega k^2\delta_1^2(1-\delta_2+\delta_3)-2v_gk^3-\omega k^2$.
Finally, the third harmonic modes ($m=3$) and ($l=1$) and with the help of Eq. \eqref{1eq:18}-\eqref{1eq:40},
give a set of equations, which can be reduced to the following NLSE:
\begin{eqnarray}
&&\hspace*{-1.3cm}i\frac{\partial\Phi}{\partial\tau}+P\frac{\partial^2\Phi}{\partial\xi^2}+Q\mid\Phi\mid^2\Phi=0,
\label{1eq:38}
\end{eqnarray}
where $\Phi=\phi_1^{(1)}$ for simplicity. In Eq. \eqref{1eq:38}, $P$ is the dispersion coefficient which can be written as
\begin{eqnarray}
&&\hspace*{-1.3cm}P=\frac{3v_g(v_g k-\omega)}{2\omega k},
\nonumber\
\end{eqnarray}
and also $Q$ is the nonlinear coefficient which can be written as
\begin{eqnarray}
&&\hspace*{-1.3cm}Q=\frac{\omega^3}{2k^2[1+\delta_1(1-\delta_2+\delta_2)]}\times\Bigg[-\frac{2k^3(H_5+H_{10})}{\omega^3}
\nonumber\\
&&\hspace*{-0.50cm}+2H_2(H_8+H_{13})+3H_3-\frac{k^2(H_4+H_9)}{\omega^2}\nonumber\\
&&\hspace*{-0.50cm}-\frac{\delta_1 k^2(1-\delta_2+\delta_2)(H_6+H_{11})}{\omega^2}\nonumber\\
&&\hspace*{-0.50cm}-\frac{2\delta_1k^3(1-\delta_2+\delta_2)(H_7+H_{12})}{\omega^3}\Bigg].
\nonumber\
\end{eqnarray}
It may be noted here that both $P$ and $Q$ are function of various
plasma parameters such as  $\alpha$, $\delta_1$, $\delta_2$, $\delta_3$, $\delta_4$ and $k$. So, all the plasma parameters are used to maintain
the nonlinearity and the dispersion properties of the PIPM.
\section{Modulational instability}
\label{1sec:Modulational instability}
The stable and unstable regions of the IAWs are organized by the sign of the dispersion ($P$)
and nonlinear ($Q$) coefficients of the standard NLSE \eqref{1eq:30}.
The stability of IAWs in four components PIPM is governed by the sign of $P$
and $Q$ \cite{Kourakis2003,Kourakis2005,Fedele2002,C2,C3,C5,Guo2013a,Guo2013b,Guo2014,Labany2015,Tantawy2013}.
When $P$ and $Q$ have same sign (i.e., $P/Q>0$), the evolution of the IAWs amplitude is modulationally
unstable in presence of the external perturbations. On the other hand, when $P$ and $Q$ have
opposite sign (i.e., $P/Q<0$), the IAWs are modulationally stable.
The plot of $P/Q$ against $k$ yields stable and unstable regions for the IAWs.
The point, at which transition of $P/Q$ curve intersects with $k$-axis, is known as threshold
or critical wave number $k$ ($=k_c$) \cite{Kourakis2003,Kourakis2005,Fedele2002,Guo2013a,Guo2013b,Guo2014,Labany2015,Tantawy2013}.

We have numerically analyzed the behaviour of $P/Q$ with respect to $k$ for different values
of $\delta_1$ and $\delta_3$ in Figs. \ref{1Fig:F1} and \ref{1Fig:F2}, respectively. It can
be seen  from Fig. \ref{1Fig:F1} that (a) the modulationally stable region corresponding to the
criteria (i.e., $P/Q<0$) as well as modulationally unstable region corresponding to the criteria (i.e., $P/Q>0$)
can be found from this figure; (b) the $k_c$ decreases with an increase in the value of $\delta_1$; (c) the mass of the
negative ions causes to increase the value of $k_c$ while the mass of the positive ions causes to decrease the value of $k_c$
for a constant value of negative and positive ions charge state (via $\delta_1$).
Figure \ref{1Fig:F2} demonstrates the presence of the Maxwellian inertialess positrons
reduces the $k_c$ as well as allows to generate IARWs associated with IAWs within the
condition $P/Q>0$ in a PIPM for a small value of $k$ when the number density as well as
the charge state of the positive ion remain constant (via $\delta_3$).
On the other hand, modulationally stable region (i.e., $P/Q<0$) as well as
$k_c$ enhances with increasing the value of number density and charge state of the positive ion
for a constant value of positron population.
\section{Rogue waves}
\label{1sec:Rogue waves}
The NLSE \eqref{1eq:38} has a variety of solutions,
among them there is a hierarchy of rational solution that are
localized in both the $\xi$ and $\tau$ variables. Each solution of the
hierarchy represents a unique event in space and time, as it
increases its amplitude quickly along each variable, reaching
its maximum value and finally decays, just as quickly as it
appeared \cite{Ankiewiez2009a}. Thus, these waves were nicknamed ``waves that
appear from nowhere and disappear without a trace \cite{Ankiewiez2009b}''.
The first-order rational solution of NLSE \eqref{1eq:38} is given as \cite{Guo2013a,Guo2013b,Guo2014,Labany2015,Tantawy2013,Ankiewiez2009a,Ankiewiez2009b,Akhmediev2009a,Akhmediev2009b,Yan2010}
\begin{eqnarray}
&&\hspace*{-1.3cm}\Phi_1 (\xi, \tau)=\sqrt{\frac{2P}{Q}}\Big[\frac{4+16 i\tau P}{1+4 \xi^2 + 16\tau^2 P^2}-1\Big] \mbox{exp} (2i\tau P).
\label{1eq:39}
\end{eqnarray}
The nonlinear superposition of the two or more first-order RWs gives rise higher-order RWs and form a higher amplitude more complicated nonlinear structure.
The second-order rational solution is expressed as \cite{Guo2013a,Guo2013b,Guo2014,Labany2015,Tantawy2013,Ankiewiez2009a,Ankiewiez2009b,Akhmediev2009a,Akhmediev2009b,Yan2010,Akhmediev2009a,Ankiewiez2009b}
\begin{eqnarray}
&&\hspace*{-1.3cm}\Phi_2 (\xi, \tau)=\sqrt{\frac{P}{Q}}\Big[1+\frac{G_2(\xi,\tau)+iM_2(\xi,\tau)}{D_2(\xi,\tau)}\Big] \mbox{exp} (i\tau P),
\label{1eq:40}
\end{eqnarray}
where
\begin{eqnarray}
&&\hspace*{-1.3cm}G_2(\xi,\tau)=\frac{3}{8}-6(P\xi\tau)^2-10(P\tau)^4-\frac{3\xi^2}{2}-9(P\tau)^2-\frac{\xi^4}{2},
\nonumber\\
&&\hspace*{-1.3cm}M_2(\xi,\tau)=-P\tau\Big[\xi^4+4(P\xi\tau)^2+4(P\tau)^4-3\xi^2+2(P\tau)^2-\frac{15}{4}\Big],
\nonumber\\
&&\hspace*{-1.3cm}D_2(\xi,\tau)=\frac{\xi^6}{12}+\frac{\xi^4(P\tau)^2}{2}+\xi^2(P\tau)^4+\frac{2(P\tau)^6}{3}+\frac{\xi^4}{8}
\nonumber\\
&&\hspace*{-0.0cm}+\frac{9(P\tau)^4}{2}-\frac{3(P\xi\tau)^2}{2}+\frac{9\xi^2}{16}+\frac{33(P\tau)^2}{8}+\frac{3}{32}.
\nonumber\
\end{eqnarray}
The solutions \eqref{1eq:39} and \eqref{1eq:40} represent the profile of the first-order and second-order RWs
within the modulationally unstable region, which concentrate a significant amount of energy into a relatively small area.
\begin{figure}[t!]
\centering
\includegraphics[width=80mm]{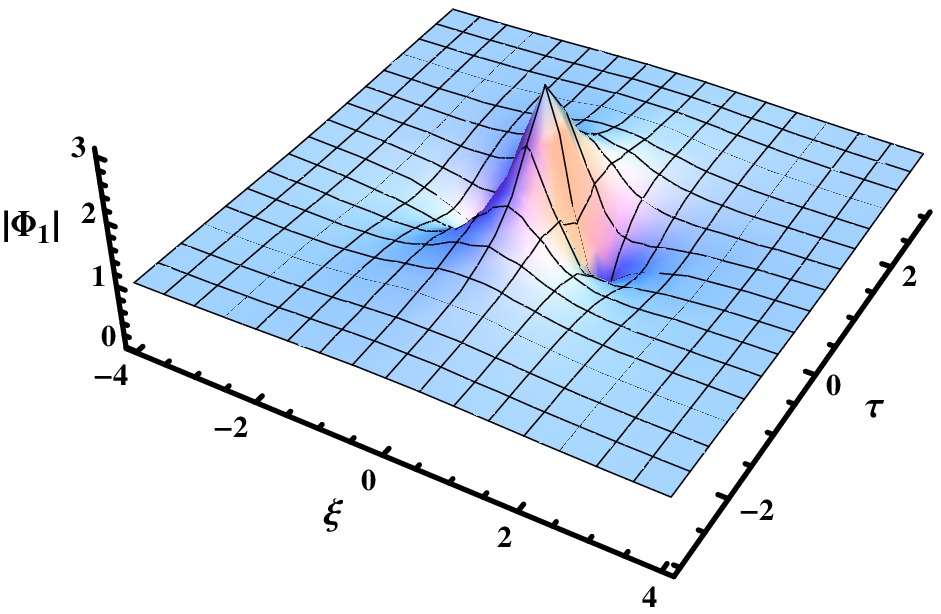}

 \Large{(a)}

\vspace{0.5cm}

\includegraphics[width=80mm]{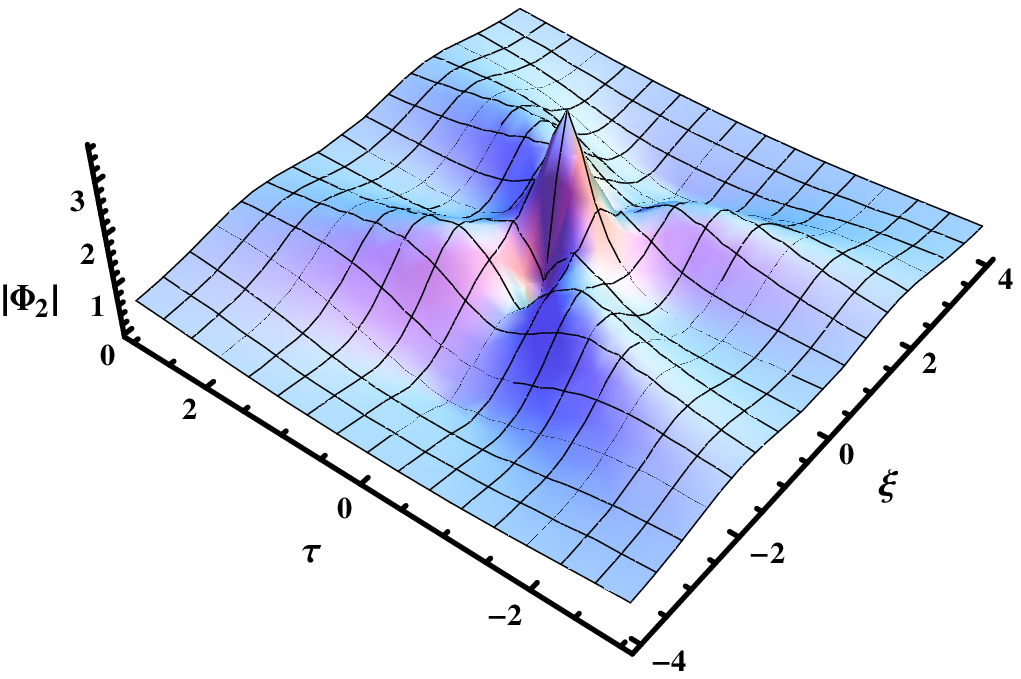}

\Large{(b)}
\caption{Profile of the (a) first-order rational solution and  (b) second-order rational solution.}
 \label{1Fig:F6}
\end{figure}
\begin{figure}
\centering
\includegraphics[width=85mm]{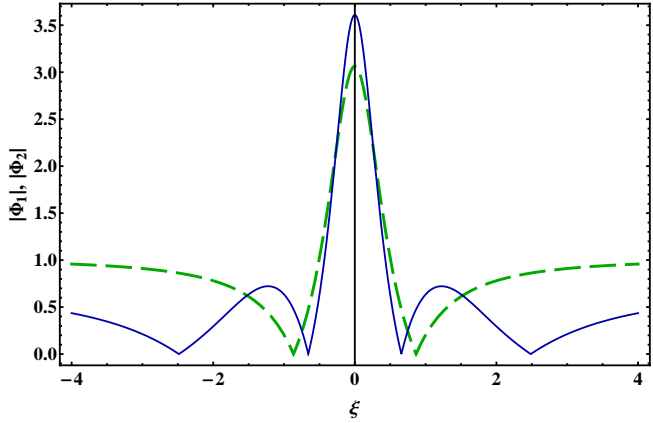}
\caption{Plot of first-order (dashed green curve) and second-order (solid blue curve) rational solutions of NLSE \eqref{1eq:38} at $\tau=0$.}
\label{1Fig:F7}
\end{figure}
The properties of the first-order rational solution [by using Eq. \eqref{1eq:39}] of NLSE \eqref{1eq:38} in a PIPM can be
observed in Figs. \ref{1Fig:F3}-\ref{1Fig:F5} corresponding different values of $\alpha$, $\delta_2$,
and $\delta_4$. Figure \ref{1Fig:F3} shows the structure of the first-order IARWs according to the value of
electrons non-thermality ($\alpha$). The increase in the value of $\alpha$ does not only cause to
increase the height of the first-order IARWs associated IAWs in the modulationally unstable region
(i.e., $P/Q>0$) but also causes to increase the thickness of the first-order IARWs associated IAWs
in the modulationally unstable region (i.e., $P/Q>0$).

The effects of the ratio of the number density of electron to positive ion as well as the charge
state of the positive ion (via $\delta_2$) can be seen from Fig. \ref{1Fig:F4},
and it is obvious from this figure that (a) the increase in the value of $\delta_4$ causes to
decrease the height of the first-order IARWs in the modulationally unstable region (i.e., $P/Q>0$)
as well as causes to decrease the thickness of the first-order IARWs in the modulationally unstable region (i.e., $P/Q>0$);
(b) physically, the presence of the excess number of electrons reduces the nonlinearity of
the plasma medium as well as the height and thickness of the first-order IARWs while the presence of the
excess number of positive ions increases the nonlinearity of the PIPM as well as the height and
thickness of the first-order IARWs when the charge state of the positive ion remains constant (via $\delta_2$).

The nature of the first-order IARWs with the variation of $\delta_4$ can be observed from Fig. \ref{1Fig:F5}
which clearly indicates that (a) the height and thickness of the first-order IARWs associated with
IAWs in the modulationally unstable region (i.e., $P/Q>0$) decreases with increasing the value of
$\delta_4$; (b) the temperature of the positron in the modulationally unstable region of IAWs
enhances the nonlinearity of the plasma medium as well as the height and thickness of the first-order IARWs while
the temperature of the electron in the modulationally unstable region of IAWs
reduces the nonlinearity of the plasma medium as well as the height and thickness of the first-order IARWs.

The space and time evolution of the first-order and second-order rational solutions of the NLSE \eqref{1eq:38} can be observed from
Figs. \ref{1Fig:F6}(a) and \ref{1Fig:F6}(b), respectively. Figure \ref{1Fig:F7} indicates the first-order and second-order solution
at $\tau=0$, and it is clear form this figure that (a) the second-order rational solution has double structures compared with  first-order rational solution;
(b) the height of the second-order rational solution is always greater than the height of the first-order rational solution; (c) the potential profile of the second-order
rational solution becomes more spiky (i.e., the taller height and narrower width) than the first-order rational solution; (d) the second (first) order rational solution
has four (two) zeros symmetrically located on the $\xi$-axis; (e) the second (first) order rational solution
has three (one) local maxima.

The existence of highly energetic rogue waves has already been confirmed
experimentally \cite{Chabchoub2011,Chabchoub2012,Bailung2011,Kibler2010}
and theoretically \cite{Shalini2015}. The second-order rational solution was experimentally observed by
Chabchoub \textit{et al.} \cite{Chabchoub2012} in a ``Water Wave Tank'',
and the experimental result regarding the amplification is a nice agreement with the theoretical result.
Bailung \textit{et al.} \cite{Bailung2011} demonstrated an experiment in a multi-component plasma medium to observe RWs, and
found a slowly amplitude modulated perturbation undergoes self modulation and gives rise to a high amplitude localized pulse.
Rogue waves also observed in fiber optics \cite{Kibler2010}.
\section{Conclusion}
\label{1sec:Conclusion}
The amplitude modulation of IAWs has been theoretically investigated in an unmagnetized four components PIPM
consisting of inertial positively and negatively charged ions as well as inertialess non-thermal electrons and iso-thermal positrons. A
NLSE, which governs the MI of IAWs and the formation of electrostatic first-order and second-order RWs in PIPM, is
derived by using the RPM. To conclude, we hope that our results may be useful in understanding the
nonlinear phenomena (viz., MI of IAWs and IARWs, etc.) in space PIPM (viz., cometary comae \cite{Sabry2009,Jannat2015,Jannat2016} and
Earth's ionosphere \cite{El-Labany2012,Massey1976,El-Tantawy2013,Abdelwahed2016,Elwakil2010})
and laboratory PIPM (viz., plasma processing reactor \cite{Gottscho1986} and neutral beam sources \cite{Bacal1979}, etc.).
\section*{Acknowledgements}
M Hassan is thankful to the Bangladesh Ministry of Science and Technology
for awarding the National Science and Technology (NST) Fellowship. A Mannan
thanks the Alexander von Humboldt Foundation for a Postdoctoral Fellowship.

\end{document}